\documentclass[runningheads]{miccai_llncs}

\usepackage{graphicx}
\usepackage[toc]{glossaries}
\usepackage[hidelinks]{hyperref}
\hypersetup{
	colorlinks=true,
	linkcolor=blue,
	filecolor=magenta,      
	urlcolor=cyan,
}

\usepackage{standalone}
\usepackage{tikz}

\usepackage{amsmath}
\usepackage{amssymb}
\usepackage{bbm}
\usepackage{eucal}
\usepackage{enumitem}
\usepackage{booktabs}
\usepackage{caption,subcaption}
\usepackage{bbm}
\usepackage{tikz}
\usepackage{adjustbox}
\usepackage{lipsum} 
\usepackage[ruled,linesnumbered]{algorithm2e}

\DeclareMathOperator*{\softmax}{\textsc{SoftMax}}

\DeclareMathAlphabet\mathbfcal{OMS}{cmsy}{b}{n}

\usepackage{tcolorbox}
\usepackage{nicefrac,xfrac}

\newcommand{\figref}[1]{Figure~\ref{#1}}
\newcommand{\tableref}[1]{Table~\ref{#1}}
\renewcommand{\eqref}[1]{Equation~\ref{#1}}

\newcommand{\mycomment}[1]{}

\usepackage{array}
\usepackage{multirow}
\newcolumntype{L}[1]{>{\raggedright\let\newline\\\arraybackslash\hspace{0pt}}m{#1}}
\newcolumntype{C}[1]{>{\centering\let\newline\\\arraybackslash\hspace{0pt}}m{#1}}
\newcolumntype{R}[1]{>{\raggedleft\let\newline\\\arraybackslash\hspace{0pt}}m{#1}}

\makeglossaries

\newacronym{drs}{DRS}{Diabetic Retinopathy Screening}
\newacronym{vqa}{VQA}{Visual Question Answering}
\newacronym{slam}{SLAM}{Simultaneous Localization and Mapping}
\newacronym[plural=CNNs,firstplural=Convolutional Neural Networks (CNNs)]{cnn}{CNN}{Convolutional Neural Network}

\newacronym{relu}{ReLU}{Rectified Linear Unit}
\newacronym{dl}{DL}{Deep learning}

\newacronym{rcc}{RCC}{renal cell carcinoma}
\newacronym{kits}{KiTS19}{Kidney Tumor Segmentation Challenge 2019}
\newacronym{ct}{CT}{computed tomography}
\newacronym{brats}{BraTS19}{Brain Tumors in Multimodal Magnetic Resonance Imaging Segmentation Challenge 2019}

\newacronym{hpc2n}{HPC2N}{High Performance Computer Center North}

\newacronym{dsc}{DSC}{S{\o}rensen-Dice coefficient}
\newacronym{se}{SE}{standard error}
\newacronym{sd}{SD}{standard deviation}
\newacronym{seb}{SEB}{Squeeze-and-Excitation block}

\newacronym{tunet}{TuNet}{TuNet: An End-to-end Hierarchical Tumor Segmentation using Cascaded U-Nets}

\usepackage{scalefnt}

\begin{document}
\title{End-to-End Cascaded U-Nets with a Localization Network for Kidney Tumor Segmentation}


\author{Minh H. Vu\inst{1} \and Guus Grimbergen\inst{2} \and Attila Simk{\'o}\inst{1}
\and Tufve Nyholm\inst{1} \and Tommy L\"{o}fstedt\inst{1}}

\institute{Department of Radiation Sciences, Ume{\aa} University, Ume{\aa}, Sweden \and Eindhoven University of Technology, 5612 AZ Eindhoven, the Netherlands}
\maketitle              

\setcounter{footnote}{0} 

\begin{abstract}

Kidney tumor segmentation emerges as a new frontier of computer vision in medical imaging. This is partly due to its challenging manual annotation and great medical impact. Within the scope of the \glsdesc{kits}, that is aiming at combined kidney and tumor segmentation, this work proposes a novel combination of 3D U-Nets---collectively denoted TuNet---utilizing the resulting kidney masks for the consecutive tumor segmentation. The proposed method achieves a \glsdesc{dsc} score of 0.902 for the kidney, and 0.408 for the tumor segmentation, computed from a five-fold cross-validation on the 210 patients available in the data.

\end{abstract}

\section{Introduction}
\label{sec:intro}

Kidney cancer has an annual worldwide prevalence of over 400\,000 new cases, with over 175\,000 deaths in 2018~\cite{bray2018global}. The most common type of kidney cancer is \gls{rcc}~\cite{motzer1996renal}. In Sweden, the indicidence of \gls{rcc} is 1\,125 per 100\,000 people, with a 0.75~\% risk of developing or dying from the disease~\cite{capitanio2018epidemiology}. Nephrectomy rates for \gls{rcc} and metastatic \gls{rcc} patients increased in Sweden between 2000 and 2008~\cite{wahlgren2013treatment}.  Renal tumors can present themselves in a wide variety in terms of size, location and depth~\cite{kutikov2009renal}. Interest has been developed to study the outcome of partial or radical nephrectomy with respect to tumor morphology~\cite{kopp2015analysis,mehrazin2013impact}. For this purpose, but also for reliable disease classification and treatment planning, accurate segmentation of renal tumors in imaging data is of utmost importance. However, manual annotation is a challenging and time-consuming process in practice, and one that is sensitive to errors and inaccuracies.

In general medical image computer vision, one of the most popular and promising methods in the last years has been \glspl{cnn}. \glspl{cnn} learn from data that has been previously segmented, often manually. \Glspl{cnn} have achieved state-of-the-art performance in organ and lesion detection, localization, segmentation, and classification.  However, the specific task of kidney and kidney tumor segmentation is challenging due to the aforementioned high degree of variability in tumor location and morphology. As a consequence, the amount of literature that can be found on this topic is limited.

One of the first attempts of using \glspl{cnn} on this task was by Yang \textit{et al.} (2018)~\cite{yang2018automatic} who first used an atlas-based approach to extract two regions of interest from the whole image, each containing one of the kidneys. These patches were used as input to a deep fully connected 3D \gls{cnn}. After additional post-processing with a conditional random field, an average \gls{dsc} of 0.931 for the kidney and 0.779 for the tumor were achieved. The Crossbar-Net by Yu \textit{et al.} (2019)~\cite{yu2019crossbar} used a stack of three rectangular patches
in the axial slice, making up what is denoted as a 2.5D data set. These patch shapes allowed for more spatial context to be incorporated, and were used to train two separate networks that combined their respective outcomes in a cascaded approach. This method was designed to segment only the renal tumor, and did so with an average \gls{dsc} of 0.913.

The present work is developed for the \gls{kits}~\cite{kits19}. Challenges in the field of medical image analysis invite the research community to develop novel approaches to common tasks. Often a data set is made publicly available to the participants, who use it to develop and test their algorithms. The challenge's organizers objectively evaluate and compare the submitted methods. The goal of the \gls{kits} challenge is to develop methods that automatically segment kidneys and kidney tumors in \gls{ct} images. The \gls{kits} challenge is part of the MICCAI 2019 conference.
We present here a method that exploits the property of this data set that one structure is a subregion of the other. The encompassing kidney and tumor region is first segmented using a variant of the popular U-Net~\cite{ronneberger2015u}. The resulting probability map is concatenated with the original image and used as input for a second network, similar in architecture, that predicts the tumor segmentation. Separately, a third network is trained to produce a mask of the whole region, that is fused with the predictions for the whole kidney region and the tumor region. We denote this method \gls{tunet}.

\begin{figure}[!t]
	\begin{center}
		\includegraphics[width=.8\textwidth]{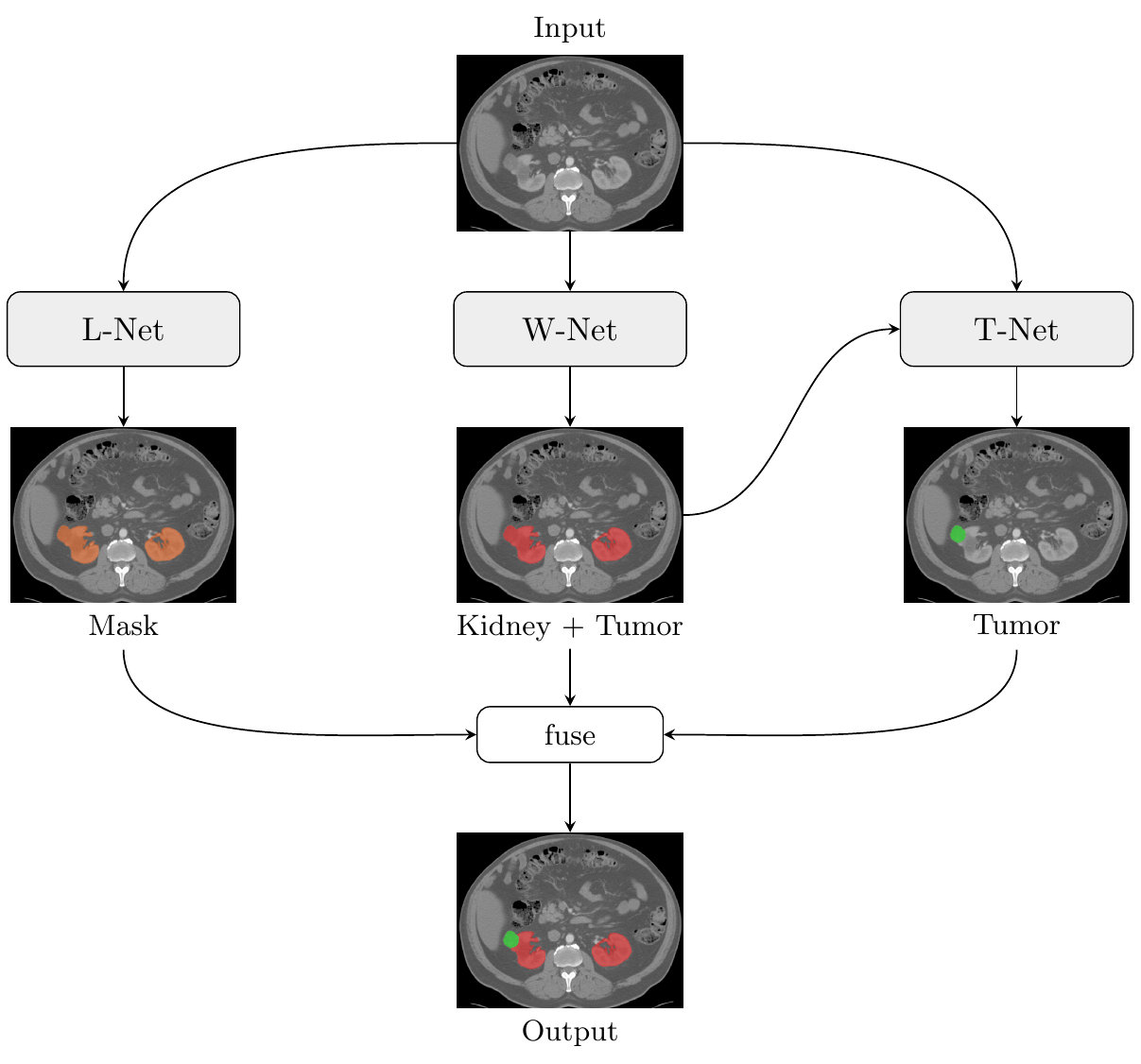}
	\end{center}
	\caption[]{The proposed Tu-Net model. The input CT image is used in a combination of U-Nets. First independently in the Localization Network (L-Net), and then in the network for whole kidney segmentation (W-Net) and finally using the kidney output probability map together with the original image in the network for the tumor segmentation (T-Net).}
	\label{fig:overview}
\end{figure}

\section{Proposed approach}
\label{sec:method}

Our method is a variation of the \gls{tunet} that we also used in the \gls{brats}. The fundamental idea for our method was based on the work of Wang \textit{et al.}~\cite{wang2017automatic} who performed multi-class brain tumor segmentation by a cascade of three networks. Their first network predicted the whole tumor region, which was used as a bounding box to generate the input for the second network. The second network in turn predicted the tumor core area from the input from the first network, and the same method was repeated for a third network that produced a prediction for the enhancing tumor core. The three outputs were finally coalesced into a single multi-label map. For solving other complex, multiple interconnected problems, with a smaller-sized network, cascaded architectures have been proposed before~\cite{kitrungrotsakul2019cascade}. The combination of two 2.5D networks have surpassed the accuracy of current state-of-the-art methods in handling automatic detection from 4D microscopic data.

For this challenge we employed a similar hierarchical architecture, since the tumor structure is always a sub-volume of the kidney + tumor structure. We first trained a network called W-Net on the whole region, and concatenated the resulting segmentation with the original image. This output was then used as the input for a second network, called T-Net, that produced a mask for the tumor only. Further, we extended the method of Wang \textit{et al.}~\cite{wang2017automatic} by separately training a localization network, denoted L-Net, that produced a crude mask of the whole region. This mask was then finally fused by multiplying with the W-Net and T-Net outputs, in order to eliminate false positives and isolated structures that lie outside the main region of interest. A summary of the proposed method is illustrated in \figref{fig:overview}.

\subsection{Localization Network}
\label{subsec:framework}

\begin{figure}[!t]
	\begin{center}
		\includegraphics[width=\textwidth]{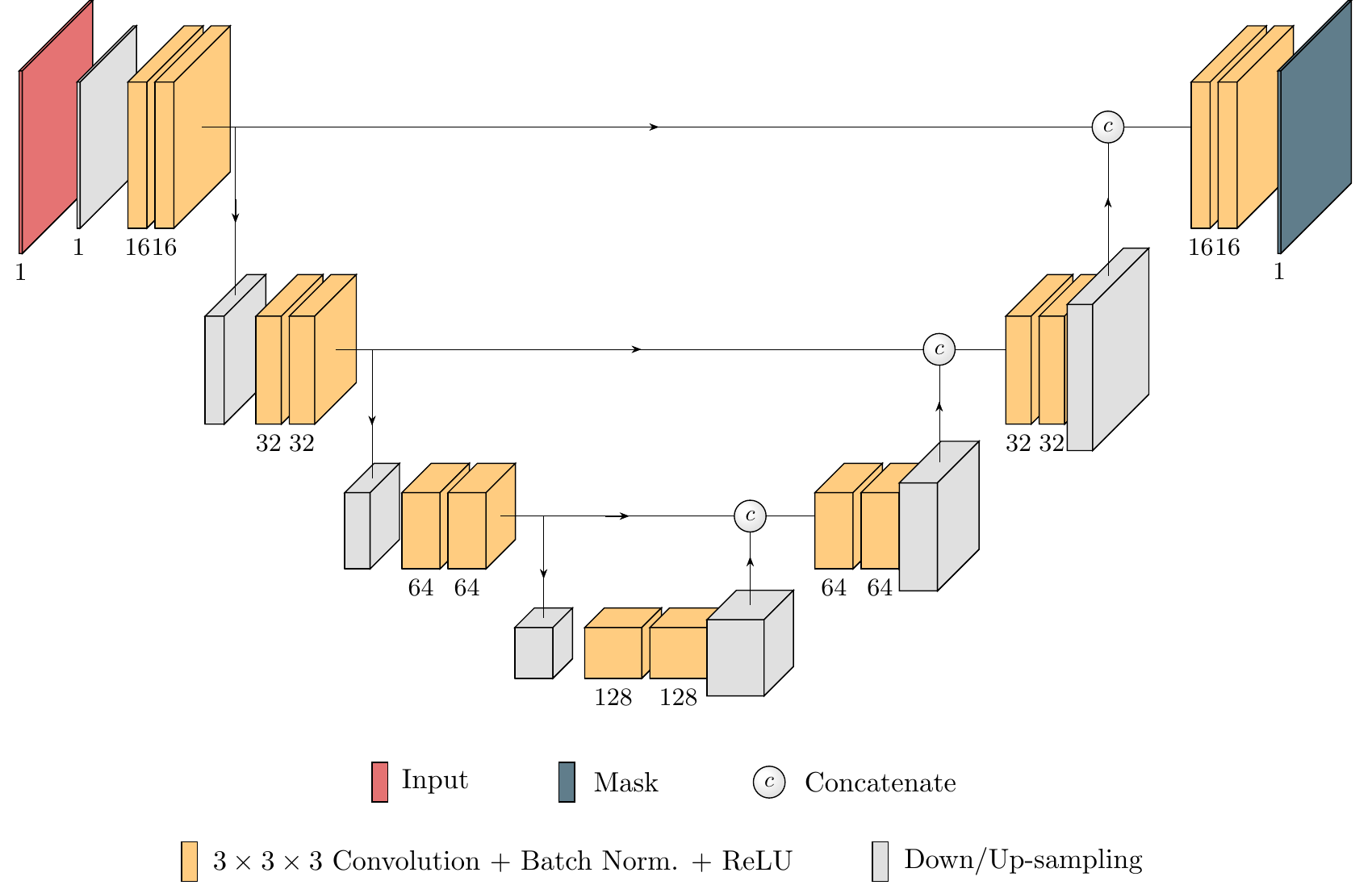}
	\end{center}
	\caption[]{The Localization Network (L-net) architecture. The figure is simplified to the single-slice case, however the convolutional layers were 3D.} 
	\label{fig:lnet}
\end{figure}

The localization network, L-Net, was an end-to-end 3D variant of the U-Net, an encoder-decoder architecture that has been proven powerful in semantic medical image segmentation~\cite{ronneberger2015u}. In order to fit the whole 3D image into the GPU memory at once, the volume was downsampled to the size of $128 \times 128 \times 64$. It was then fed into a traditional 3D U-Net to estimate roughly the region of interest of the whole (kidney + tumor) region. The L-Net is illustrated in \figref{fig:lnet}. For simplicity of illustration, it is shown with a 2D input and output.

\subsection{Cascaded Architecture}
\label{subsec:framework}

The model architectures used for the whole region and tumor segmentation had a similar U-Net architecture to the localization network. An additional feature of both the W-Net and the T-Net was that in each convolution block in the encoding path, the original input image was downsampled and concatenated with the feature map of the respective layer.
The rate of downsampling was increased according to the dimensions of that layer.

\begin{figure}[!t]
	\begin{center}
		\includegraphics[width=\textwidth]{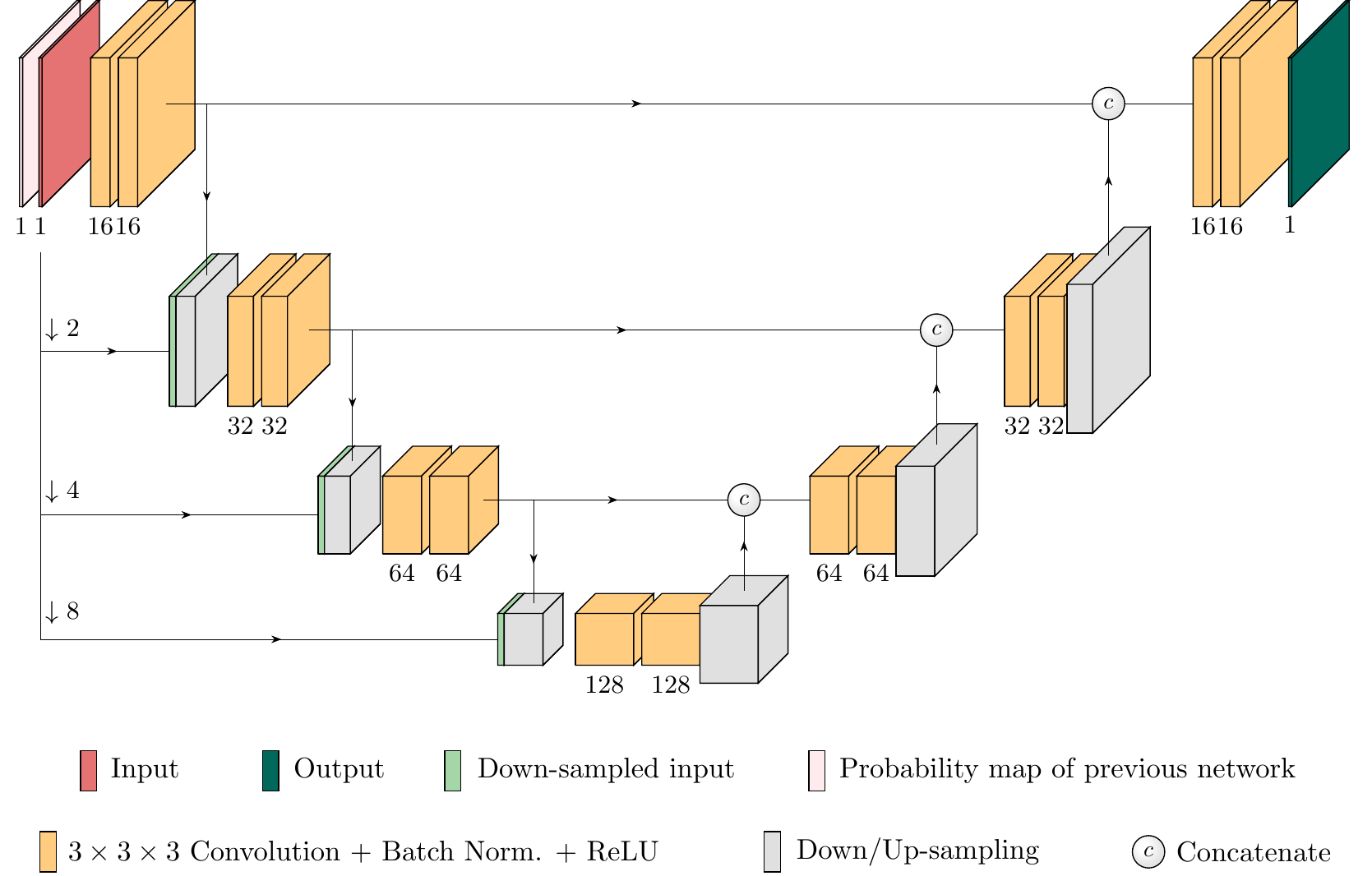}
	\end{center}
	\caption[]{The W-Net and T-Net architecture. The figure is simplified to the single-slice case, however the convolutional layers were 3D.}
	\label{fig:tunet}
\end{figure}

As mentioned before, we employed a cascaded approach. The input of the T-Net consisted of two channels: the input image, and the probability map of the previous output. An illustration of the T-Net architecture is given in \figref{fig:tunet}. The W-Net was analogous, but lacked the probability map as secondary input.

\subsection{Squeeze-and-Excitation Block}
\label{subsec:se}

After each convolution and concatenation block, we added an \gls{seb} as developed by Hu \textit{et al.} (2018)~\cite{hu2018squeeze}. Including \gls{seb} is a computationally lightweight but effective method for incorporating channel-wise interdependencies, and has shown to improve network performance by a significant margin~\cite{DBLP:journals/corr/abs-1902-01314}.


\section{Experiments}
\label{sec:experiment}

The proposed method was implemented in the Keras library\footnote{\url{https://keras.io}} using TensorFlow\footnote{\url{https://tensorflow.org}} as the backend. This research was conducted using the resources of the High Performance Computing Center North (HPC2N), where the experiments were run on NVIDIA Tesla V100 GPUs.

\subsection{Material}
The data set provided is pre- and postoperative, arterial phase abdominal \gls{ct} data from $300$ randomly selected kidney cancer patients that underwent radical nephrectomy at the University of Minnesota Medical Center between 2010 and 2018~\cite{kits19}. Medical students annotated under supervision the contours of the whole kidney including any tumors and cysts, and contours of only the tumor component excluding all kidney tissue. Afterwards, voxels with a radiodensity of less than $-30$~HU were excluded from the kidney contours, as they were most likely perinephric fat.

Of the complete data set, $210$ samples were initially released for training and validation on March 15, 2019. The remaining $90$ samples were released without ground truth on July 15. The submission deadline for the predictions on the latter set was July 29.

\subsection{Training Details}

In addition to the experiments with single patch sizes, we also additionally denoised the data using a $3 \times 3 \times 3$ median filter. This yielded a total of four different experiments.

\begin{figure}[!t] 
    \centering
    \includegraphics[width=0.8\textwidth]{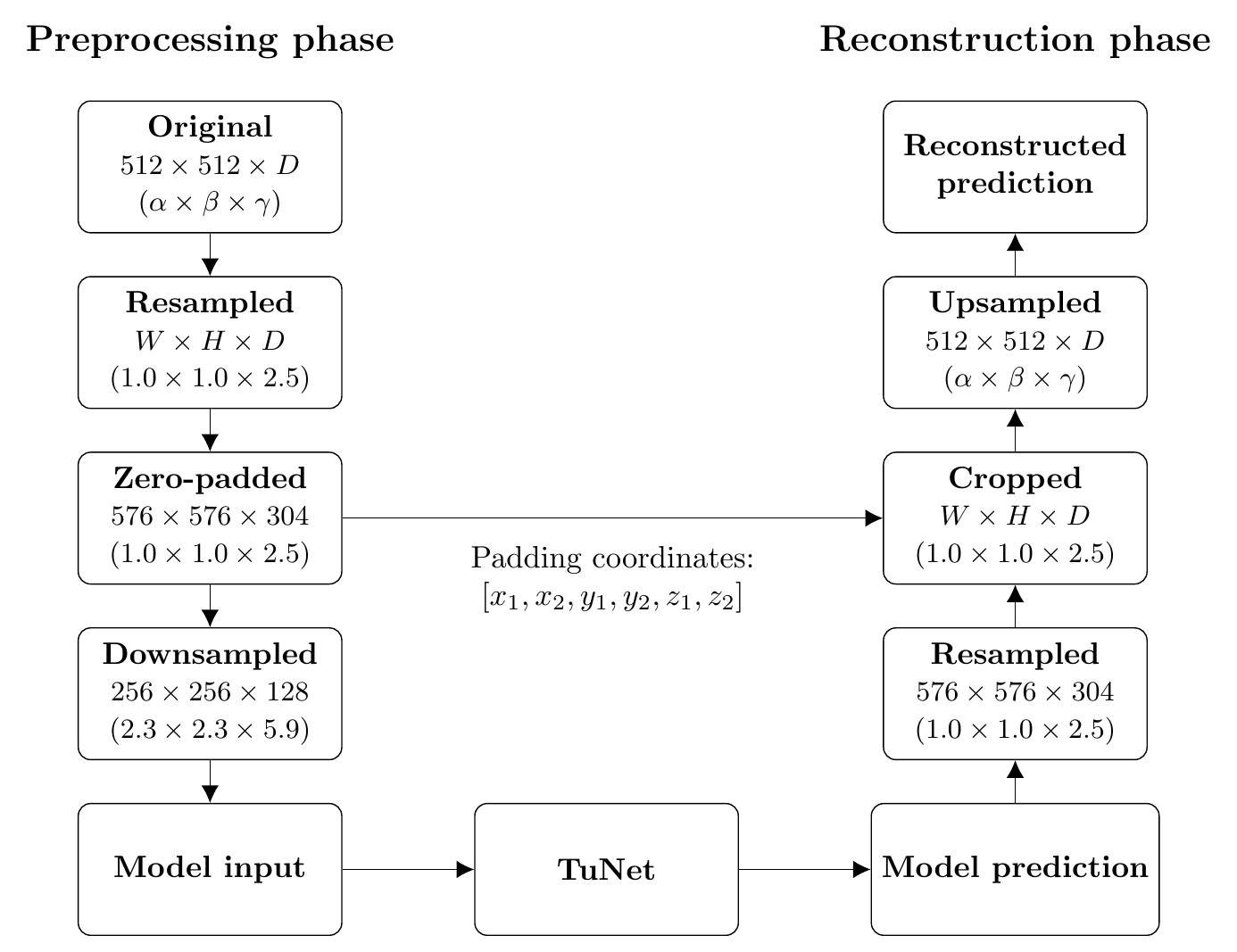}
    \caption{Preprocessing and reconstruction pipeline method as applied on the \gls{kits}
             data set. Given are the resolutions and in parentheses the voxel dimensions in mm. $W$, $H$, and $D$ each denote that the volume shape is varied in width, height, and/or depth, respectively. $\alpha$, $\beta$, and $\gamma$ each denote that the voxel spacing is varied in width, height, and/or depth, respectively. $[x_1, x_2, y_1, y_2, z_1, z_2]$ denote the number of pixels padded above and below in the $x$, $y$ and $z$ directions, respectively.}
    \label{fig:preprocess}
\end{figure}

The loss function that was used contained two terms,
\begin{equation}
    \mathcal{L} = \mathcal{L}_{\textit{whole}} + \mathcal{L}_{\textit{tumor}},
\end{equation}
where $\mathcal{L}_{\textit{whole}}$ and $\mathcal{L}_{\textit{tumor}}$ denote the soft dice loss of the whole (kidney + tumor) and tumor regions, respectively. Here, the soft dice loss for any structure was defined as
\begin{equation}
    \mathcal{L}_{dice} = \frac{-2 \sum_i u_i v_i}{\sum_i u_i + \sum_i v_i + \epsilon},
\end{equation}
where $u_i$ is the $\softmax$ output produced by either the W-Net or the T-Net, $v_i$ is a one-hot encoding of the ground truth segmentation map, and $\epsilon$ is a small constant added to avoid division by zero.

For evaluation of the segmentation performance, we employed the \gls{dsc} which is defined as
\begin{equation}
    D(X,Y)=\frac{2 |X \cap Y|}{|X| + |Y|},
\end{equation}
where $X$  and $Y$ denote the output segmentation and its corresponding ground truth, respectively. We used five-fold cross-validation, and computed the mean \gls{dsc} over the folds.

We used the Adam optimizer~\cite{kingma2014adam} with a learning rate of $1 \cdot 10^{-4}$ and momentum parameters of $\beta_1=0.9$ and $\beta_2=0.999$. We also used $L_2$ regularization (with parameter $1 \cdot 10^{-5}$), applied to the kernel weight matrix, for all convolutional layers to cope with overfitting. The activation function of the final layer was the sigmoid function. 

We trained 20 models for $200$ epochs (see \tableref{tab:cv}). To prevent overfitting, we selected a patience period by dropping the learning rate by a factor of $0.2$ if the validation loss did not improve over $6$ epochs. In addition, the training process was stopped if the validation loss did not improve after $15$ epochs. The training time was approximately $27$ hours per model.


\subsection{Patch Extraction}

We employed a patch-based approach in which we set up the experiments with a patch shape of $128 \times 128 \times 128$. To extract more patches from each image, we allowed a patch overlap of $96 \times 96 \times 96$. The size of the patch shape was large enough to contain each kidney, when centered.

\subsection{Preprocessing and Augmentation}

All images in the data set had $512 \times 512$ resolution with varying slice counts in the axial direction. The voxel sizes also varied between patients. A preprocessing pipeline (see \figref{fig:preprocess}) was set up to transform the data to uniform resolutions and voxel sizes.

First, the data was resampled to equal voxel size of $256 \times 256 \times 128$. The volumes were then zero-padded to the size of the single largest volume after resampling. In order to increase processing speed and lower the memory consumption, the volumes were thereafter downsampled to size $256 \times 256 \times 128$. As a final step, the images were normalized to zero-mean and unit-variance.

In order to augment the data set size and diversify the data, we employed simple on-the-fly data augmentation by randomly rotating the images within a range of $-1$ to $1$ degrees.

\subsection{Post-processing}

Algorithm \ref{alg:post} shows our proposed post-processing pipeline. Each 3D probability map for the whole region was binarized using a threshold $T = 0.5$, which in case of an empty mask, was lowered to $T = 0.1$.

The corresponding probability map for the tumor region was first binarized using the threshold $T = 0.5$ which was lowered based on the following three cases: If the returned mask is empty, $T$ is lowered to $0.1$. If the returned mask is not empty but it contains less than 100 pixels, the lowered threshold is $T = 0.2$, and in any other case $T$ is lowered to $0.3$ to include more positives. The optimal value for $T$ was tuned using the training data set.

The binary masks for the whole and tumor regions were then refined using a ``region-growing'' technique: given two thresholds. First, two binarized maps, $A$ and $B$,  were generated using the larger and smaller threshold, respectively. Second, the binary map, $B$, was labelled into different unconnected regions. Third, if a segmented region was connected to the binary map, $A$, it was then merged into $A$. The output of the refining step is the expanded region of $A$. 

The refined binary masks were finally merged with the output of L-Net to produce a multi-class prediction map.

\begin{algorithm}[!t]
    \SetAlgoLined
    \KwResult{Post-processed prediction map}
    set global threshold $T=0.5$\;
    binarize probability maps of whole and tumor regions based on $T$\;
    \eIf{number of non-zeros voxels of binarized whole region equals 0}{
        set $T_{whole}=0.1$\;
        binarize probability map of whole region based on $T_{whole}$\;
    }{
    set $T_{whole}=0.4$\;
    refine whole region based on $T$ and $T_{whole}$\;
    }
    
    \eIf{number of non-zeros voxels of binarized tumor region equals 0}{
        set $T_{tumor}=0.1$\;
        binarize probability map of tumor region based on $T_{tumor}$\;
        \eIf{number of non-zeros voxels of binarized tumor region equals 0}{
        set tumor equal to whole region\;
        }
        
        refine tumor region based on $T$ and $T_{tumor}$\;
    }{
    \eIf{number of non-zeros voxels of binarized tumor region smaller than 100}{
    set $T_{tumor}=0.2$\;
    refine tumor region based on $T$ and $T_{tumor}$\;
    }{
    set $T_{tumor}=0.3$\;
    refine tumor region based on $T$ and $T_{tumor}$\;
    }
    }
    merge whole and tumor regions to produce multi-class prediction map\;
    multiply prediction map with output of L-Net to generate final output\;
    \caption{Post-process}
    \label{alg:post}
\end{algorithm}

\subsection{Ensembling}

The median denoising and \gls{seb} led to four experiments: the baseline method (TuNet), the baseline on median denoised data (TuNet + median), the baseline with Squeeze-and-Excitation blocks (TuNet + SEB), and the baseline with Squeeze-and-Excitation blocks on median denoised data (TuNet + SEB + median). As a fifth experiment, an average ensemble technique was implemented, using the output probability maps of the above mentioned models, as
\begin{equation} \label{eqn:ensemble_avg}
    p_{k} = \frac{1}{M} \sum_{m=1}^{M} f_{k},
\end{equation}
where $p_k \in \mathbb{R}^{|K|}$ and $f_k \in \mathbb{R}^{|K|}$ denote the final probability of label $k$ and the probability of label $k$ generated by model $m$ at an arbitrary voxel, respectively. Here, $K$ is the set of tumor labels, \textit{i.e.} $k=1,2$.

Each model was trained five independent times, their ensemble network (Ensemble) is therefore using a total of 20 models. As a last experiment, this ensemble model was used with an extra post-processing step (Ensemble + post-process) described in Algorithm~\ref{alg:post}.

\section{Results and Discussion}
\label{sec:results}

Evaluations are collected below for the results of the vanilla form of U-Nets~\cite{ronneberger2015u}, for TuNet, its three variations: TuNet + median, TuNet + SEB, and TuNet + SEB + median, and the ensemble models: Ensemble and Ensemble + post-process. They underwent evaluations based on the DSC metric, the best variation was then also evaluated visually, see \figref{fig:qualitative}.

\subsection{Quantitative analysis}

\tableref{tab:cv} shows the mean \gls{dsc} score and \gls{sd} computed from the five-folds of cross-validation on the training set. The baseline method produced acceptable scores for the whole region segmentation, while the significant impairment of results for tumor segmentation are important to point out. They are possibly due to the scarcity of tumors in the data set and because of their spatial inconsistency, which the proposed network could not invalidate.

The \gls{seb} variations improved the \gls{dsc} scores over the baseline method for the whole region while impairing the tumor segmentation, leading to only a slight effect on the mean score, whether using median denoised data or not.

However, the baseline method using median denoised data significantly improved the \gls{dsc} scores for both the whole and the tumor region, and although the \gls{sd} for the tumor masks also increased, the improvements and effects of median denoising are still clear and important.

The Ensemble model
reached the same mean score as the best of the other variations, \textit{i.e.} TuNet + median. The post-processing step significantly improved the Ensemble model, making it the best-performing model of all models.


\subsection{Qualitative analysis}

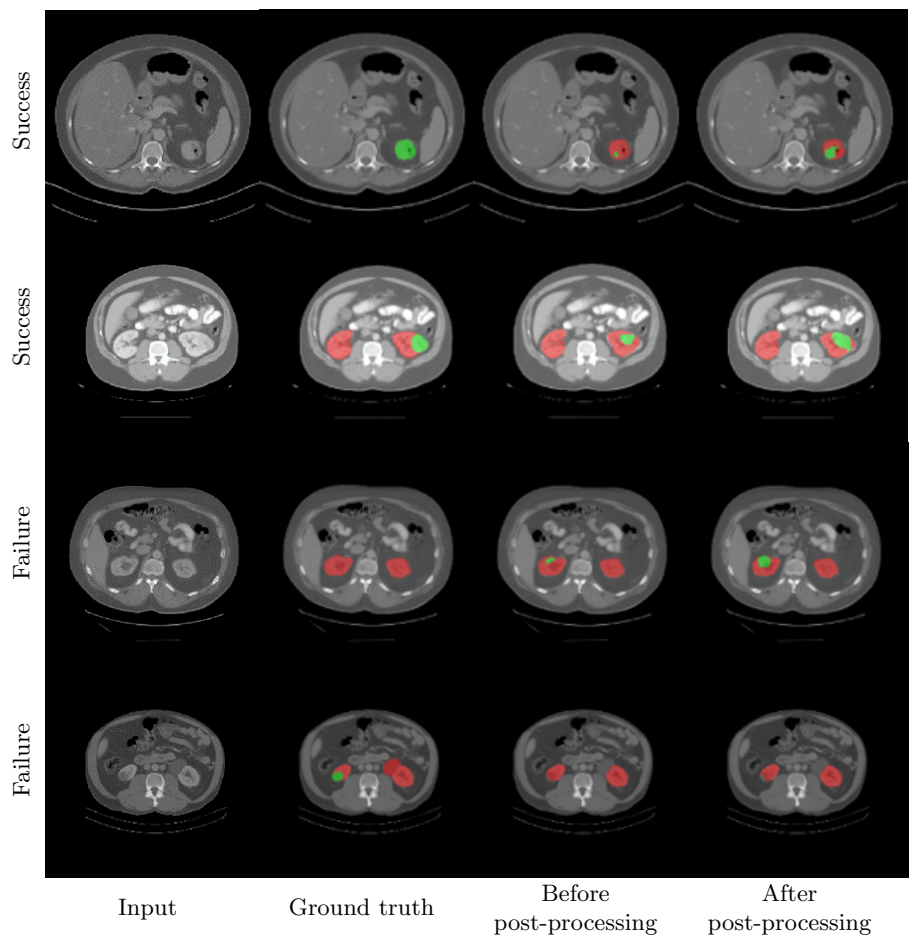
\begin{figure}[!t]
    \def\subfigsize{0.24\textwidth}
    \def\subfigverlabelsize{0.24\textwidth}
    \def\hspaceminus{-0.3cm}
    \def\vpaceminus{-0.3cm}
    \begin{center}
        \begin{tikzpicture}
            \node [rotate=90, text width=\subfigverlabelsize, text centered] {Success};
        \end{tikzpicture}
        \includegraphics[width=\subfigsize]{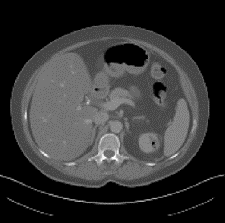}
        \hspace{\hspaceminus}
        \includegraphics[width=\subfigsize]{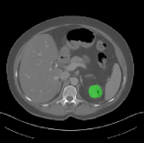}
        \hspace{\hspaceminus}
        \includegraphics[width=\subfigsize]{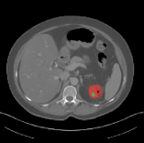}
        \hspace{\hspaceminus}
        \includegraphics[width=\subfigsize]{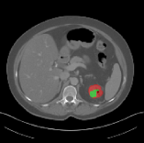} \\
        \vspace{\vpaceminus}
        \begin{tikzpicture}
            \node [rotate=90, text width=\subfigverlabelsize, text centered] {Success};
        \end{tikzpicture}
        \includegraphics[width=\subfigsize]{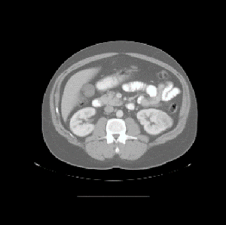}
        \hspace{\hspaceminus}
        \includegraphics[width=\subfigsize]{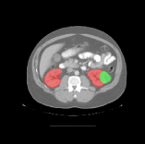}
        \hspace{\hspaceminus}
        \includegraphics[width=\subfigsize]{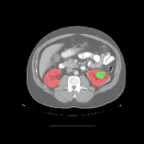}
        \hspace{\hspaceminus}
        \includegraphics[width=\subfigsize]{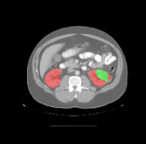} \\
        \vspace{\vpaceminus}
        \begin{tikzpicture}
            \node [rotate=90, text width=\subfigverlabelsize, text centered] {Failure};
        \end{tikzpicture}
        \includegraphics[width=\subfigsize]{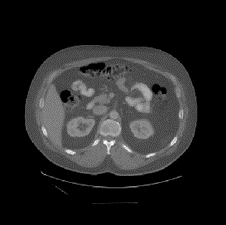}
        \hspace{\hspaceminus}
        \includegraphics[width=\subfigsize]{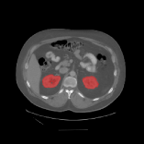}
        \hspace{\hspaceminus}
        \includegraphics[width=\subfigsize]{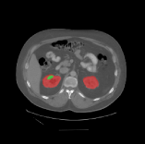}
        \hspace{\hspaceminus}
        \includegraphics[width=\subfigsize]{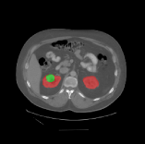} \\
        \vspace{\vpaceminus}
        \begin{tikzpicture}
            \node [rotate=90, text width=\subfigverlabelsize, text centered] {Failure};
        \end{tikzpicture}
        \includegraphics[width=\subfigsize]{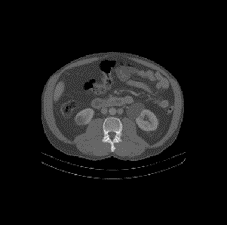}
        \hspace{\hspaceminus}
        \includegraphics[width=\subfigsize]{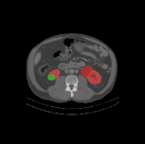}
        \hspace{\hspaceminus}
        \includegraphics[width=\subfigsize]{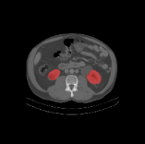}
        \hspace{\hspaceminus}
        \includegraphics[width=\subfigsize]{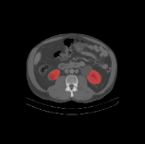} \\
        \begin{tikzpicture}
            \node [rotate=90, text width=\subfigverlabelsize/10, text centered] {~};
        \end{tikzpicture}
        \captionsetup[subfigure]{labelformat=empty, justification=centering}
        \begin{subfigure}{\subfigsize}
            \caption{Input}
        \end{subfigure}
        \hspace{\hspaceminus}
        \begin{subfigure}{\subfigsize}
            \caption{Ground truth}
        \end{subfigure}
        \hspace{\hspaceminus}
        \begin{subfigure}{\subfigsize}
            \caption{Before post-processing}
        \end{subfigure}
        \hspace{\hspaceminus}
        \begin{subfigure}{\subfigsize}
            \caption{After post-processing}
        \end{subfigure} \\
        \vspace{\vpaceminus}
    \end{center}
    \caption{A comparison of the ground truth masks and the results of Ensemble +
             post-process. The four examples show the input CT images, the ground truth masks and the results of the proposed method. For both cases the masks show the kidney+tumor region in red while the tumors are in green.}
    \label{fig:qualitative}
\end{figure}

\figref{fig:qualitative} demonstrates the inter-patient diversity of the structures in this data set, where the results are produced using the Ensemble + post-process model. The renal tumors can be present in left, right, or both kidneys, and their size in the images can vary from several cubic centimeters to multiple times the volume of the kidney it is attached to. The intratumoral radiodensity can also be either very homogeneous or very heterogeneous. This results in a challenging segmentation task. The morphology of the kidneys is however relatively comparable across patients and can therefore be annotated with a fairly consistent and acceptable accuracy.


\begin{table}[!t]
    \def\width{2.5 cm}
    \def\widthdesc{4 cm}
    \caption{Mean \gls{dsc} and \gls{sd} computed from the five-folds of cross-validation on the training set. \gls{tunet}, median, \gls{seb} denote the baseline, with median denoising and with \glsdesc{seb}, respectively. Note that the results were reported on the resampled output (size of $256 \times 256 \times 128$, see \figref{fig:preprocess}).
    }
    \centering
    \begin{adjustbox}{max width=\textwidth}
    \begin{tabular}{l c C{\width} C{\width} C{\width}}
    \toprule
    \multirow{2}{*}{}       && \multicolumn{3}{c}{S{\o}rensen-Dice coefficient}   \\
    Model                   && Kidney + Tumor& Tumor         & Mean          \\
    \cmidrule{1-1}\cmidrule{3-5}
    U-Net 2D                && 0.776 (0.009)  & 0.174 (0.036)  & 0.475 (0.017) \\
    U-Net 2D + median       && 0.781 (0.016)  & 0.172 (0.030)  & 0.476 (0.020) \\
    U-Net 3D                && 0.793 (0.038)  & 0.154 (0.019)  & 0.474 (0.016) \\
    U-Net 3D + median       && 0.809 (0.075)  & 0.192 (0.056)  & 0.501 (0.065) \\
    \cmidrule{1-1}\cmidrule{3-5}
    TuNet                   && 0.857 (0.029)  & 0.308 (0.034)  & 0.582 (0.030) \\
    TuNet + median          && 0.878 (0.010)  & 0.346 (0.053)  & 0.612 (0.031) \\
    TuNet + SEB             && 0.868 (0.020)  & 0.302 (0.035)  & 0.585 (0.023) \\
    TuNet + SEB + median    && 0.864 (0.033)  & 0.292 (0.017)  & 0.578 (0.023) \\
    \cmidrule{1-1}\cmidrule{3-5}
    Ensemble                && 0.894 (0.014)  & 0.331 (0.025)  & 0.612 (0.026) \\
    Ensemble + post-process && \textbf{0.902 (0.014)}  & \textbf{0.408 (0.031)}  & \textbf{0.655 (0.019)} \\
    \bottomrule                  
    \end{tabular}
    \end{adjustbox}
    \label{tab:cv}
\end{table}

\section{Conclusion}
\label{sec:conclusion}

The \gls{kits} Challenge is clearly justified given the difficulty of segmenting the images in the data set. Initially the provided data set was in itself anticipated to be insufficient for excelling results for this work. For training the proposed TuNet architecture, we used only a simple on-the-fly data augmentation, and still produced reasonable \gls{dsc} scores for both the whole region and the tumor region compared to their corresponding models using a simple U-Net in both 2D and 3D.

Two expansions of the TuNet method were evaluated. Using \gls{seb} has only a slight effect on the mean scores, while median denoising the data leads to significant improvements. An ensemble model using these variations with a simple post-processing step outperformed all other models and reached impressive \gls{dsc} scores.

\bibliographystyle{splncs04}
\bibliography{bib}

\end{document}